\documentclass[11pt,twoside]{article}

\usepackage{asp2006}
\usepackage{epsf}
\usepackage{graphicx}
\usepackage{lscape}
\usepackage{amssymb}
\usepackage{natbib}

\def\apjl{ApJ}          
\def\apjs{ApJS}         
\def\aap{A\&A}          
\def\mnras{MNRAS}       

\newcommand{\Sauron}{\texttt{SAURON}}

\def\deg{^{\circ}}

\def\spose#1{\hbox to 0pt{#1\hss}}
\def\lta{\mathrel{\spose{\lower 3pt\hbox{$\sim$}}
    \raise 2.0pt\hbox{$<$}}}
\def\gta{\mathrel{\spose{\lower 3pt\hbox{$\sim$}}
    \raise 2.0pt\hbox{$>$}}}

\begin{document}

\title{Towards a new classification of early-type galaxies: an integral-field view}   

\author{Jes\'us Falc\'on-Barroso,\altaffilmark{1} 
        Roland Bacon,\altaffilmark{2} 
        Michele Cappellari,\altaffilmark{3}\\
	Roger~L. Davies,\altaffilmark{3} 
        P.~Tim de Zeeuw,\altaffilmark{4} 
        Eric Emsellem,\altaffilmark{2}\\
        Davor Krajnovi\'{c},\altaffilmark{3} 
	Harald Kuntschner,\altaffilmark{5} 
        Richard~M. McDermid,\altaffilmark{4} 
        Reynier~F. Peletier,\altaffilmark{6}
	Marc Sarzi,\altaffilmark{7} and 
        Glenn van de Ven\altaffilmark{8}}   

\altaffiltext{1}{European Space and Technology Centre, Keplerlaan 1, 2200~AG Noordwijk, The Netherlands}
\altaffiltext{2}{Universit\'e de Lyon 1, CRAL, Observatoire de Lyon, 9 av. Charles Andr\'e, 69230 Saint-Genis Laval, France}
\altaffiltext{3}{Sub-Department of Astrophysics, University of Oxford, Denys Wilkinson Building, Keble Road, Oxford OX1~3RH, United Kingdom}
\altaffiltext{4}{Sterrewacht Leiden, Universiteit Leiden, Postbus 9513, 2300~RA Leiden, The Netherlands}
\altaffiltext{5}{Space Telescope European Coordinating Facility, European Southern  Observatory, Karl-Schwarzschild-Str.~2, 85748 Garching, Germany}
\altaffiltext{6}{Kapteyn Astronomical Institute, University of Groningen, NL-9700 AV Groningen, The Netherlands}
\altaffiltext{7}{Centre for Astrophysics Research, University of Hertfordshire,  Hatfield, Herts AL10~9AB, United Kingdom}
\altaffiltext{8}{Institute for Advanced Study, Einstein Drive, Princeton, NJ~08540, USA}

\begin{abstract} 
In this proceeding we make use of the two-dimensional stellar kinematics of a
representative sample of E and S0 galaxies obtained with the \Sauron\
integral-field spectrograph to reveal that early-type galaxies appear in two
broad flavours, depending on whether they exhibit clear large-scale rotation or
not. We measure the level of rotation via a new parameter ($\lambda_R$) and use
it as a basis for a new kinematic classification that separates early-type 
galaxies into slow and fast rotators. With the aid of broad-band imaging we
will reinforce this finding by comparing our kinematic results to the photometric 
properties of these two classes.
\end{abstract}

\section{Introduction}
The origins of the morphological classification of galaxies date back from 
early work by Jeans in 1929. The latter addition of the S0 galaxies as a new
class by \citet{Hubble36} resulted in the "tunning-fork" diagram that we know
and use today. The Hubble classification is a continuous sequence between
ellipticals (E), lenticulars (S0) and spirals (S) with the S0s occupying the
transition region. Elliptical and lenticular galaxies are usually  gathered
into the so-called early-type category given the large number of global
photometric properties they share \citep{RC3}.\looseness-1

More recently, with the advent of CCD imaging, there has been several attempts
to  revise the current scheme to introduce a more physical description of the
objects, and therefore go beyond a purely descriptive tool. With this goal in
mind \citet[][hereafter KB96]{KB96} updated the Hubble sequence by sorting 
ellipticals in terms of photometric quantities, used as a proxy for the
importance of rotation. They used the disciness or boxiness of the isophotes to
define refined types: E(d) galaxies (for discy ellipticals) making the link
between S0s and E(b) galaxies (for boxy ellipticals). This extension of the
Hubble types has the merit of upgrading our view of Es and S0s via some easily
accessible observable parameter that, at the same time,  includes some physics
into the sorting criteria. It did, however, use a photometric indicator as an
attempt to quantify the dynamical state of the galaxy, which may deem
unreliable.

In this contribution, we revisit the early-type galaxy classifications above 
using the available full two-dimensional kinematic information coming from the
unique data set obtained with the \Sauron\ integral-field spectrograph. A more 
compeling study of the kinematics for this sample with a discussion on the
kinematical classification presented here can be found in \citet{Emsellem07}.\looseness-2

\begin{figure} 
\centering 
\includegraphics[width=0.99\linewidth]{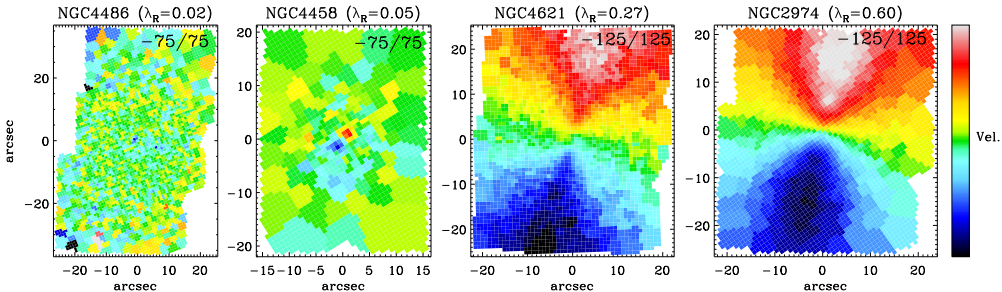} 
\caption{Stellar velocity fields for 4 elliptical galaxies in the \Sauron\
sample of early-type galaxies. The maps are sorted from left to right by
increasing value of the $\lambda_{R}$ parameter. The first two (on the left)
galaxies belong to the Slow rotator class, whereas the last two are Fast
rotators. Note the different kinematical substructures despite belonging to the
same elliptical class. Numbers on the top right corner of each map indicate the
velocity cuts applied to the data, which have been adjusted as to properly
emphasise the observed velocity structures\looseness-2}
\label{fig:maps} 
\end{figure}

\section{A kinematical classification of galaxies}

The velocity fields for our sample of 48 E and S0 galaxies, basis for this
study, were presented in \citet{Emsellem+04}. They revealed a large variety
of kinematical structures including decoupled cores, velocity twists,
misalignments, cylindrical or disc-like rotation. A close examination of the
maps suggests that  early-type galaxies come in two broad flavours: one which
exhibits a clear large-scale and rather regular rotation pattern, and another
which shows perturbed velocity structures (e.g. strong velocity twists) or
central kinematically decoupled components with little rotation in the outer
regions. Here, we illustrate these features in Figure~\ref{fig:maps} using 4
elliptical galaxies (NGC\,4486, NGC\,4458, NGC\,4621, NGC\,2974).
From the  figure it is easy to recognize the large differences in kinematical
substructure  between the galaxies despite belonging to the same E
morphological class.\looseness-1

\begin{figure} 
\centering 
\includegraphics[width=0.99\linewidth]{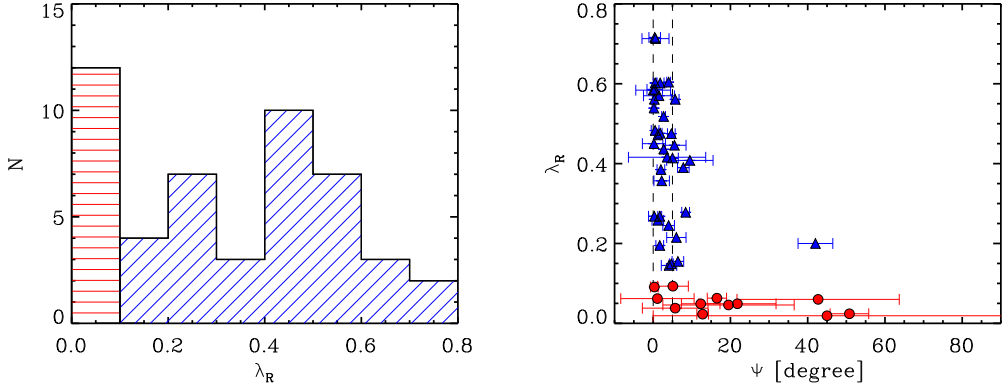} 
\caption{(left) Distribution of galaxies as a function of $\lambda_{R}$: the horizontally striped
bar indicates the bin for slow rotators, whereas diagonally striped histogram corresponds to fast
rotators. (right) $\lambda_{R}$ versus the kinematic misalignment $\Psi$
between  the global photometric major-axis and the kinematic axis within the
\Sauron\ field.  Slow rotators are represented by filled circles, fast rotators by
solid triangles.  The vertical dashed lines correspond to $\Psi = 5\deg$. Nearly
all fast rotators have small $\Psi$ values ($< 10\deg$), the only exception
being NGC\,474, the photometry of which is perturbed by the presence of
irregular shells.  This contrast with slow rotators which show significantly
non-zero $\Psi$ values.\looseness-2}
\label{fig:misc} 
\end{figure}

In order to build a robust classification based on the observed kinematics we
need  a simple measurable parameter which quantifies the {\em global}
dynamical  state of a galaxy, and that can be applied to all galaxies in our
sample. The ideal tool would be a physical parameter which captures the spatial  
information included in the kinematic maps. Since we wish to assess the level of 
rotation in galaxies, this parameter should follow the nature of the classic 
$V / \sigma$: ordered versus random motion (see \citealt{Emsellem07} for a detailed
explanation of the pitfalls of the $V / \sigma$ parameter as a reliable classification
parameter). Following this idea, we have defined a new quantity $\lambda_{R}$:

\begin{displaymath} 
\label{eq:sumLambda} 
\lambda_R \equiv \frac{\langle R \, \left| V \right| \rangle }{\langle R \, \sqrt{V^2 + \sigma^2} \rangle}\, ,
\end{displaymath}

that measures the amount of specific (projected) angular momentum from the
velocity maps. The parameter has been defined such that is insensitive to small
features in the maps, and therefore provides a robust measurement of the global
rotation. As we go from galaxies with low to high $\lambda_{R}$ values (from
left to right in Fig.~\ref{fig:maps}), the overall velocity amplitude naturally
tends to increase. More importantly, there seems to be a {\em qualitative}
change in the observed stellar velocity structures. Rotators with $\lambda_{R}
<0.1$ exhibit low stellar mean velocities at large radii, with very perturbed
stellar kinematics and {\it all} have large-scale kinematically decoupled
components.  In Figure~\ref{fig:misc} {\it(left)} we show the distribution of
our galaxies as a function of $\lambda_{R}$. There are 36 fast rotators and 12
slow rotators (75\% and 25\% of the total sample), their median $\lambda_{R}$
values being $\sim 0.44$ and $0.05$ respectively. Within the class of slow
rotators, three galaxies have $\lambda_{R}$ significantly  below 0.03 (their
mean stellar velocity maps being consistent with  zero rotation everywhere).
These are among the brightest galaxies of our sample (NGC\,4486, NGC\,4374 and
NGC\,5846).\looseness-2

\section{Comparing photometry and kinematics: misalignments and twists}
In order to assess whether our kinematical classification based on the
$\lambda_{R}$  parameter is solid, it is desirable that the structural
differences seen in the  kinematics are also reflected in the photometry. In
Figure~\ref{fig:misc}{\it (right)} we show that slow and fast rotators display
clear differences in the global alignment between their photometric and
kinematic major axes ($\Psi \equiv | \mbox{PA}_{phot} - \mbox{PA}_{kin}|$). 
The figure illustrates that all fast rotators, except one, have misalignments
$\Psi$ below $10\deg$. The only exception is NGC\,474 which is an interacting
galaxy with well-know irregular shells  \citep{Turnbull+99}.  In fact, the few
galaxies which have $5\deg < \Psi < 10\deg$  (NGC\,3377, NGC\,3384, NGC\,4382,
NGC\,4477, NGC\,7332) are most probably barred. In contrast, more than half of
all slow rotators have $\Psi > 10\deg$, and none of these exhibit any hint of a
bar. This difference in the misalignment values of slow and fast rotators
cannot be entirely due to the effect of inclination, mainly because even the
roundest fast rotators do not exhibit large misalignment values.

Independent evidence in support for our kinematical classification comes from
the  observed velocity twists in the \Sauron\ kinematic maps: only 6 galaxies
out of 48 exhibit strong velocity twists larger than $30\deg$ outside the inner
3\arcsec, with 3 out of these 6 having large-scale counter-rotating stellar
components (NGC\,3414, 3608, and 4550).  All these galaxies are in fact slow
rotators.  This implies that only fast rotators have a relatively well defined
(apparent) kinematic major-axis, which in addition is roughly aligned with the
photometric major-axis. As emphasised above slow and fast rotators exhibit
qualitatively but also quantitatively different kinematical properties. This
suggests that slow rotators are not just scale-down versions of fast rotators.

\acknowledgements 
JFB would like to thank the organizers for a very fruitful meeting and
wishes to congratulate J. Beckman for his outstanding contribution to 
astronomy in particular, but science in general, in the last 40 years.\looseness-2

\end{document}